\documentclass{article}
\usepackage{siunitx}
\usepackage{spconf,amsmath,graphicx, url, multirow, makecell, xcolor, array}
\usepackage{hyperref}

\usepackage{enumitem}
\setlist{nosep, leftmargin=14pt}
\usepackage{bbding}

\usepackage{mwe} 

\definecolor{chosen_blue}{RGB}{4,50,255}
\definecolor{chosen_red}{RGB}{255,0,0}
\definecolor{chosen_yellow}{RGB}{255,192,0}
\definecolor{chosen_green}{RGB}{112,173,71}
\definecolor{chosen_orange}{RGB}{255,135,47}

\title{Gyral-Sulcal-Net: An Integrated Network Representation of Brain Folding Patterns}
\name{%
\parbox[t]{\textwidth}{\centering
Chao Cao$^{1}$\sthanks{Co-first authors.}, 
Tong Chen$^{1}$\footnotemark[1], 
Nan Zhao$^{2}$\footnotemark[1], 
Minheng Chen$^{1}$, 
Michael Qu$^{3}$, 
Zeyu Zhang$^{1}$, 
Xiao Shi$^{1}$,\\
Xiang Li$^{4}$, 
Tianming Liu$^{5}$, 
Lu Zhang$^{2}$\sthanks{Corresponding author: lz50@iu.edu.}
}}

\address{\textsuperscript{1} Computer Science and Engineering, The University of Texas at Arlington, Arlington, TX, USA
    \\
    \textsuperscript{2} Department of Computer Science, Indiana University Indianapolis, Indianapolis, IN, USA 
    \\
    \textsuperscript{3} Mission San Jose High School, Fremont, CA, USA
    \\
    \textsuperscript{4} Department of Radiology, Massachusetts General Hospital and Harvard Medical School, MA, USA
    \\
    \textsuperscript{5} School of Computing, The University of Georgia, Athens, GA, USA
}

\begin{document}
%
\maketitle
\begin{abstract}
Our brain functions as a complex communication network, and studying it from a network perspective offers valuable insights into its organizational principles and links to cognitive functions and brain disorders. However, most current network studies typically use brain regions as nodes, often overlooking the intricate folding patterns of finer-scale anatomical landmarks within these regions. In this study, we introduce a novel approach to integrate the brain's two primary folding patterns -- gyri and sulci -- into a unified network termed the Gyral-Sulcal-Net (GS-Net), in which three different types of finer-scale landmarks have been successfully identified. We evaluated the proposed GS-Net across multiple datasets, comprising over 1,600 brain scans, spanning different age groups (from 34 gestational weeks to elderly adults) and cohorts (healthy brains and those with pathological conditions). The experimental results demonstrate that the GS-Net can effectively integrate and represent diverse cortical folding patterns from a network perspective. More importantly, this approach offers a promising way for integrating different folding patterns into a unified anatomical brain network, alongside structural and functional networks, providing a comprehensive framework for studying brain networks.
\end{abstract}

\begin{keywords}
Brain Network, Folding Pattern, Gyri and Sulci, Finer-Scale Landmarks.
\end{keywords}
\section{Introduction}
\label{sec:intro}

Our brain is a communication network, where neural elements are interconnected with each other, creating a structural foundation~\cite{doi:10.1126/science.1238411, Seguin2023}. Upon this structural substrate, signaling and information transmission permeate across various levels and spatial scales of brain activity, governing the function, cognition, and behavior of a normal brain \cite{Petersen2015}. Furthermore, studies on brain disorders have revealed that pathological perturbations of the brain often spread via axonal pathways (network connections) to influence other regions, rather than confining to a single locus \cite{Fornito2015}. This indicates that disease propagation patterns are constrained by the intricate and highly organized topology of the brain’s underlying network architecture. Therefore, adopting a network perspective to study the brain has shown significant potential for uncovering organizational principles within the brain and their connections to cognitive procedures and brain disorders.

Brain network study starts with the identification of network nodes as interacting units and their interconnections as edges. In neuroimaging studies, various parcellation methods \cite{ARSLAN20185}, including those based on anatomical landmarks (e.g., AAL), cytoarchitectonic information (e.g., Brodmann areas), and connectivity patterns, are employed to subdivide the brain into a set of brain regions, which are utilized as network nodes. Advanced imaging techniques, such as Magnetic Resonance Imaging (MRI), are employed to construct macro-scale networks based on these ROIs, including the structural network inferred from diffusion MRI (dMRI) and the functional network estimated using resting-state functional MRI (rs-fMRI) \cite{doi:10.1073/pnas.0811168106,https://doi.org/10.1002/hbm.22933}. Structural network allows estimation of the physical connections, while the functional network elucidates putative functional connections. 
Brain network studies have made significant breakthroughs in understanding the fundamental organizational principles of the normal brain \cite{ZHANG2022102463} and brain diseases \cite{Bassett9239}. However, current research still lacks sufficient investigation into two critical aspects. Firstly, most existing brain networks utilize brain regions as nodes, which often cover large areas at a relatively coarse scale. The folding pattern of finer-grained anatomical landmarks within these brain regions tends to be overlooked, such as sulci fundi (deepest loci of sulci) and gyri peaks (highest loci in gyri) \cite{Thecerebralsulciandgyri}. Accumulating evidences suggest that these landmarks confer specific structural, functional, and cognitive patterns \cite{10.1093/cercor/bhac537} and are under a stronger genetic or heritability influence than other cortical regions \cite{10.1093/cercor/bhac537,IM2019881}. Therefore, establishing a finer-scale brain network based on these landmarks is significant and would serve as a valuable complement to the region-based brain network. Secondly, existing studies primarily employ structural and functional connectivity to construct brain networks, and anatomical folding patterns cannot be effectively integrated into such frameworks. Although certain studies have investigated the unique structural-functional patterns of specific folding patterns and their changes in disease development \cite{KIKINIS19947,10.3389/fpsyt.2022.1033918}, these methods often independently analyze particular folding patterns. They fail to effectively integrate different folding patterns within a unified brain network, alongside the brain structural and functional networks, to comprehensively study the brain from a network perspective.
To address these problems, in this study, we introduced a novel framework to integrate the brain’s two primary folding patterns, gyri and sulci, into a unified network called Gyral-Sulcal-Net (GS-Net), while also automatically identifying three important landmarks -- gyri conjunctions, sulci conjunctions, and sulci-gyri conjunctions. We evaluated the proposed GS-Net using multiple datasets comprising 1,600+ brain scans. These datasets cover various age groups ranging from 34 gestational weeks to elderly adults, and different cohorts including healthy brains and brains with pathological conditions. The generated GS-Net can accurately describe the folding pattern of the brain from a network perspective.

\section{Method}
\label{sec:method}

\subsection{Method Overview}
\label{ssec:method_overview}
\begin{figure}[htb]

\begin{minipage}[b]{1.0\linewidth}
  \centering
  \centerline{\includegraphics[width=8.5cm]{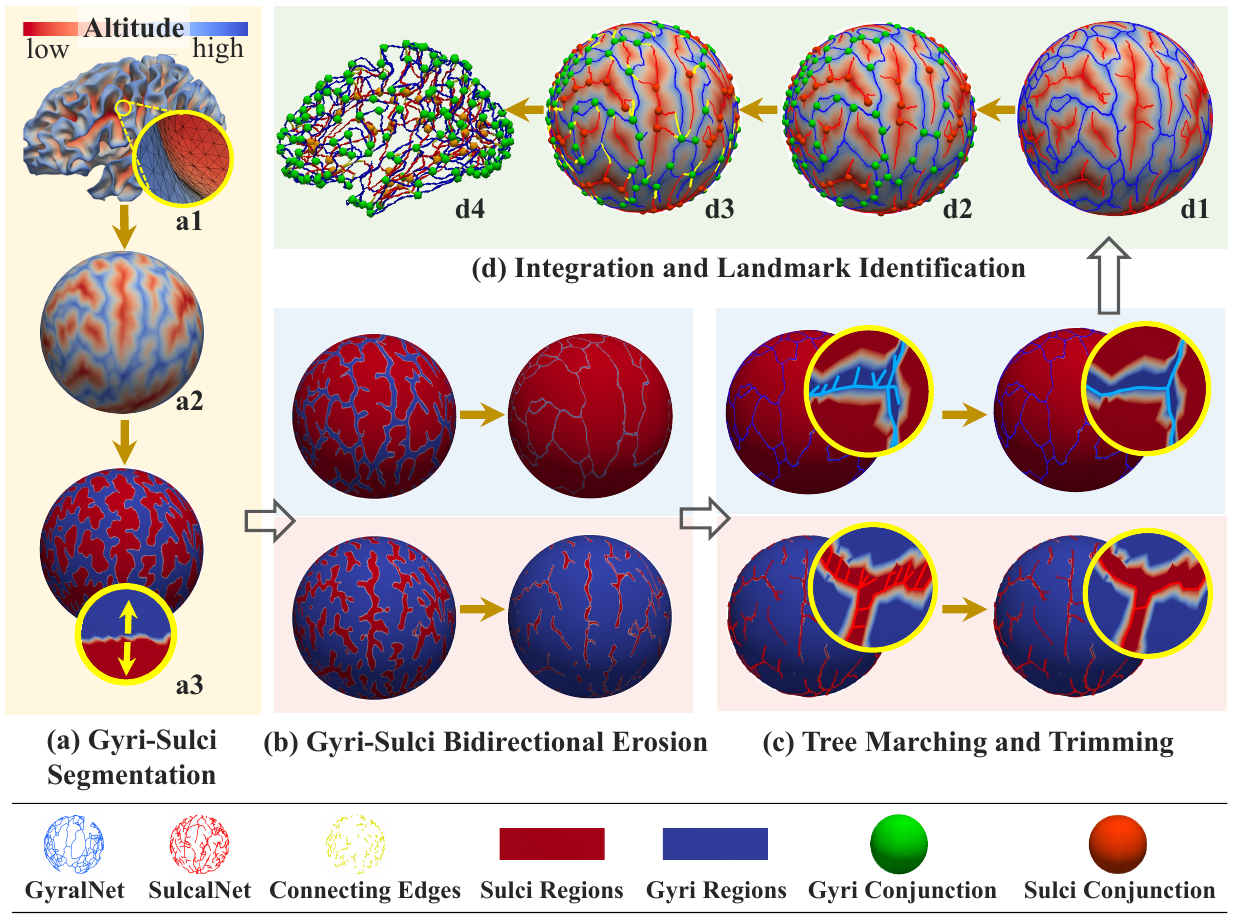}}
  \caption{The pipeline of the proposed GS-Net framework.}
  \label{fig1}
\end{minipage}
\end{figure}

The pipeline of the proposed GS-Net is illustrated in Fig.~\ref{fig1}, comprising four main steps: a) Gyri-Sulci segmentation. This step aims to partition the entire cerebral cortex into two distinct regions: sulci and gyri. b) Gyri-sulci bidirectional erosion. Following segmentation, this process involves erosion along the boundary between gyri and sulci regions on both sides until reaching the main skeletons of both gyri and sulci regions. c) Tree marching and trimming. In this step, a tree marching and trimming algorithm is utilized to connect the remaining skeleton regions into two independent networks: GyralNet and SulcalNet. d) Integration and landmark identification. In this step, GyralNet and SulcalNet are integrated into a unified network, and three different types of landmarks are identified. 

\subsection{Gyri-Sulci Segmentation}
\label{ssec:segmentation}
In this work, T1-weighted MRI was used to reconstruct the white matter surface using a meshing algorithm \cite{FISCHL2012774}. The generated surface is represented as a triangle mesh in 3D space (Fig.~\ref{fig1} (a1)). Then gyral altitudes were computed for individual vertices and projected onto the mesh surface. Gyral altitude represents the displacement of a vertex on the surface from an imaginary “mid-surface” positioned between gyri and sulci to its original location. This mid-surface is chosen to ensure that the average displacement of all surface vertices from their original positions equals zero \cite{https://doi.org/10.1002/(SICI)1097-0193(1999)8:4<272::AID-HBM10>3.0.CO;2-4}. To provide a clearer representation of the distribution and pattern of gyri and sulci regions on the main surface, we inflated the surface into a sphere (Fig.~\ref{fig1} (a2)) via the FreeSurfer package \cite{FISCHL2012774}. Subsequently, based on positive and negative altitude values, the sphere surface was partitioned into gyri (\textcolor{blue}{blue}) and sulci (\textcolor{red}{red}) regions (Fig. ~\ref{fig1} (a3)).

\subsection{Gyri-Sulci Bidirectional Erosion}
\label{ssec:erosion}
This study aims to establish network representations of cortical gyri and sulci, with edges representing the primary branches of each gyrus and sulcus, known as gyri crests and sulci valleys. These features are crucial for delineating the essential morphology of the gyri and sulci. To achieve this, we first apply the watershed algorithm to extract the gyri crests and sulci valleys across the entire region. The watershed algorithm is widely used in image processing. It applies a transformation to grayscale images, treating them as topographic maps where the brightness of each point corresponds to its height. This transformation identifies lines that traverse the summits of ridges, aligning with our objective of locating gyri crests and sulci valleys on the cortical surface.

The core steps of the watershed algorithm involve initializing a water source at each regional minimum within the relief map. Water levels are then progressively raised to flood the entire relief from these sources, constructing barriers where flows from different water sources converge. In our study, the gyri and sulci regions can effectively serve as initial water sources for one another. For example, when identifying gyri crests (Fig.~\ref{fig1} (b), upper block), the sulci region can be used as the initial water source. By continuously extending the sulci region along the borders of the gyri and sulci, the gyri region experiences erosion until the different sulci regions nearly merge. To ensure precise identification of gyri crests, the erosion process is carefully controlled by the altitude of the gyri. The threshold for gyri altitude is increased only after all regions below the current threshold have been eroded, allowing further erosion in areas with higher gyri altitudes. Once the erosion process is complete, the remaining regions correspond to the gyri crests. The identification of sulci valleys follows a similar process (Fig.~\ref{fig1} (b), lower block).

\subsection{Tree Marching and Trimming}
\label{ssec:marching_and_trimming}
After segmenting the gyri crests and sulci valleys, we employed tree marching and trimming methods (Fig.~\ref{fig1} (c)) to construct networks upon gyri crests and sulci valleys. Taking the gyri crest regions as an example, the tree marching process begins with the vertex at the highest gyri altitude and progressively connects other vertices in descending order until all vertices within the gyri crests are linked. During this process, all vertices in the crest area are connected, resulting in some redundant branches extending from the skeleton to the boundary. We then perform trimming to eliminate these extraneous branches. Since the remaining gyri crests form narrow regions, the redundant branches that extend beyond the main structure are relatively short and can be easily identified. Following the trimming, the main skeleton is integrated into a network that delineates the primary morphology of the gyri regions, which we refer to as GyralNet. Similarly, SulcalNet is constructed for the sulci valley regions.

\subsection{Integration and Landmark Identification}
\label{ssec:landmark_identification}
The convex gyri and concave sulci together form the folded surface of the cerebral cortex. Sulci are surrounded by, and intersect with, gyri, which leads to the fragmentation of sulcal regions, while most gyri remain connected. As a result, SulcalNet appears more like a collection of subnetworks and free edges compared to GyralNet. The challenge lies in linking these separated elements without imposing rigid connections that could disrupt the brain's natural patterns.  Previous study \cite{Thecerebralsulciandgyri} indicates that the presence of a free sulcal extremity implies the existence of a fold connecting different gyri or sectors within the same gyrus. Thus, extending along a free sulcal extremity will inevitably lead to either a junction point in GyralNet, where multiple edges converge, or to a midpoint along an edge. To avoid introducing anatomical inaccuracies by adding new network nodes, we connected only those free sulci whose extremities extended directly to a junction node within GyralNet. Our findings show that over 95\% of free sulci meet this criterion. Consequently, the majority of gyri and sulci can be integrated into a unified network, which we have named Gyral-Sulcal-Net (GS-Net). 

GS-Net contains three distinct types of nodes, each playing a key role in brain networks. The first type includes conjunctions of multiple edges in GyralNet, representing intersections of various gyri on the cerebral cortex. These nodes have been highlighted in the literature as regions with thicker cortices, higher fiber density, and significant variations in structural and functional connectivity \cite{LI20101202,10.1093/cercor/bhx227}. Based on their connections to free sulci, we further classify these nodes as GC (Gyri Conjunction) and SGC (Sulci-Gyri Conjunction) in this study. Additionally, the conjunctions of multiple edges in SulcalNet show unique structural, functional, and genetic characteristics. Existing study suggests these nodes exhibit spatial consistency across individuals during development \cite{Thecerebralsulciandgyri} and are subject to stronger genetic or heritability influences compared to other cortical regions \cite{10.1093/cercor/bhac537}. These nodes are referred to as SC (Sulci Conjunction).

\section{Results}
\label{sec:results}

\subsection{Dataset Description and Data Pre-processing}
\label{ssec:dataset_description}
We utilized three diverse datasets, comprising a total of 1,623 brain scans, to evaluate the effectiveness of our proposed methods. These datasets span various age groups, from 34 gestational weeks to elderly adults, and include both healthy subjects and patients diagnosed with Alzheimer's Disease (AD). Specifically, we used T1-weighted structural MRI scans from 1,064 young adults in the Human Connectome Project (HCP) S1200 release; 
282 elderly adults, and 200 AD patients from the ADNI dataset; 
and 27/30/20 subjects in 2-year/6-month/34-week groups from the dHCP datasets. 
The pre-processing steps included brain skull removal, tissue segmentation, and cortical surface reconstruction conducted via the FreeSurfer package \cite{FISCHL2012774}.

\setlength{\doublerulesep}{0.3mm}   

\begin{figure}[htb]
\begin{minipage}[b]{1.0\linewidth}
  \centering
  \centerline{\includegraphics[width=7.8cm]{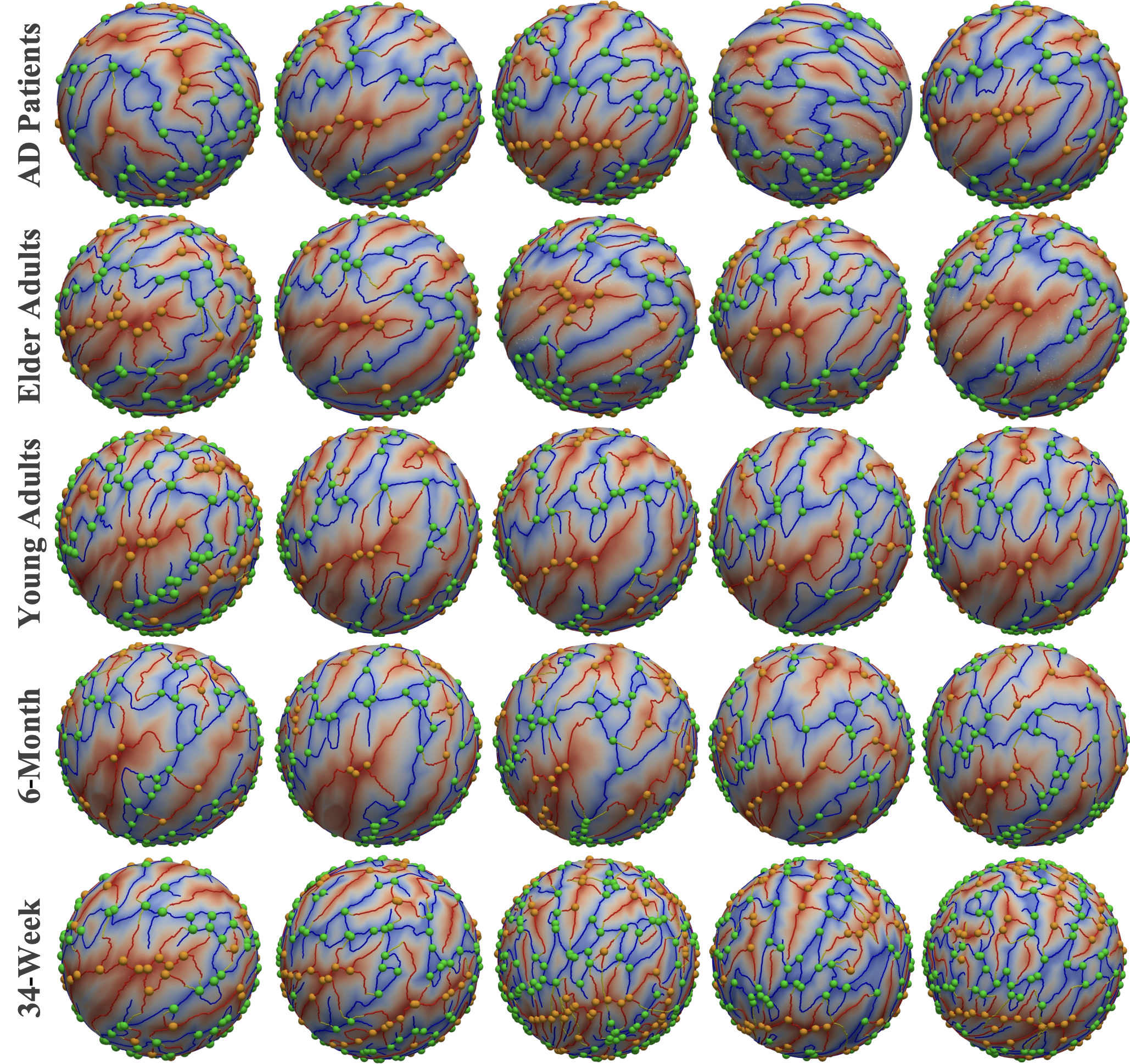}}
  \caption{Constructed GS-Net of 25 randomly selected subjects from 5 different cohorts, including AD patients, elderly adults, young adults, individuals at 6 months, and individuals at 34 weeks’ gestation.}
  \label{fig2}
\end{minipage}
\end{figure}

\subsection{Constructed GS-Net}
\label{ssec:constructed_gyral_sulcal_net}
In our experiments, we constructed GS-Net for each individual and presented the results of 25 randomly selected subjects from five different cohorts, including AD patients, elderly adults, young adults, individuals at 6 months, and individuals at 34 weeks of gestation. 
These results are illustrated in Fig.~\ref{fig2}. 
Each subject's sphere surface is color-coded by gyri altitude, depicting the specific patterns of gyri and sulci of the individual. The generated GS-Net is overlaying on the sphere surface, comprising five components: GyralNet (\textcolor{chosen_blue}{blue} lines), SulcalNet (\textcolor{chosen_red}{red} lines), edges connecting GyralNet and SulcalNet (\textcolor{chosen_yellow}{yellow} lines), GC (gyri conjunction) nodes (\textcolor{chosen_green}{green} bubbles), and SC (sulci conjunction) nodes (\textcolor{chosen_orange}{orange} bubbles). 
Notably, the deepest red and blue regions on the sphere surface represent the locations with the highest positive and negative gyri altitude, corresponding to the primary gyri crests and sulci valleys. 
Our generated GS-Net accurately covers these regions and effectively connects gyri regions and sulci regions along the folding pattern into a unified network. 
This highlights the precision and effectiveness of GS-Net.

\begin{figure}[htb]
\begin{minipage}[b]{1.0\linewidth}
  \centering
  \centerline{\includegraphics[width=8.5cm]{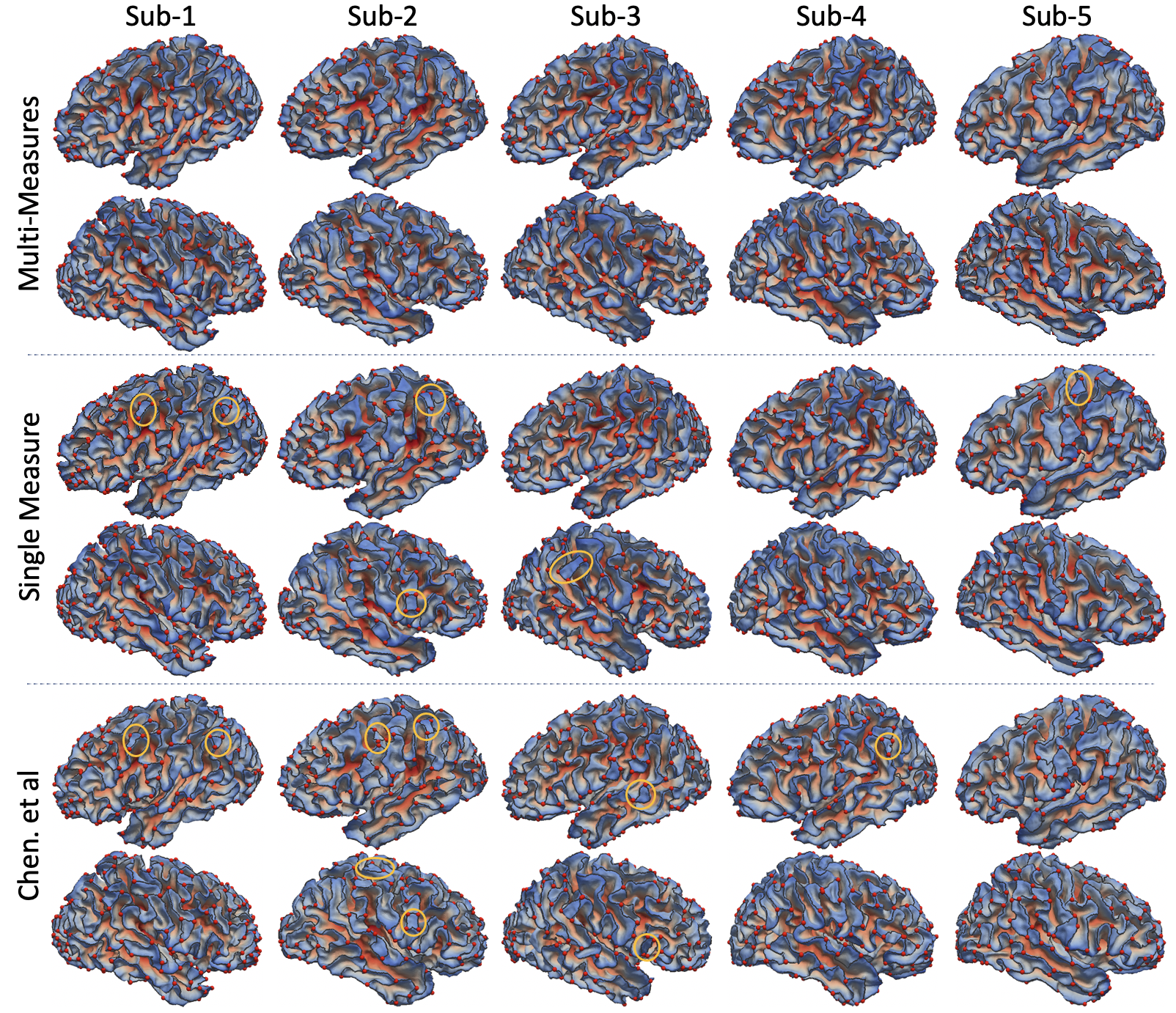}}
  \caption{Qualitative comparison across five subjects shows that both the Single-Measure sulc-based baseline and the traditional method frequently produce redundant or spurious 3HGs (yellow circles). In contrast, our Multi-Measures approach yields more accurate and consistent 3HGs.}
  \label{comparsion}
\end{minipage}
\end{figure}

\subsection{Comparison with Baseline Methods}
\label{sec:Comparison}

We compared our method with two baselines: a single-measure sulc-based approach and the method of~\cite{chen2017gyral}. 
Because cortical folding patterns are highly complex, relying on a single measure is more sensitive to noise and often leads to inaccurate 3-hinge gyrus (3HG) localization. 
As shown in Fig.~\ref{comparsion}, these baselines frequently produce redundant 3HG detections (yellow circles). 
The previous method~\cite{chen2017gyral} also uses a single measure and depends on multiple hand-crafted stopping parameters during tree construction and pruning. 
Due to large inter-subject variability, one parameter set cannot generalize well, causing good performance in some subjects but redundant or missing 3HG branches in others. 
In contrast, our approach jointly uses sulc and curv, providing richer geometric cues, and employs an adaptive termination criterion based on complete erosion that naturally identifies the population-consistent gyral crest. 
This eliminates the need for fixed thresholds and yields more accurate and stable 3HG detection across individuals.

\begin{figure}[htb]
\begin{minipage}[b]{1.0\linewidth}
  \centering
  \centerline{\includegraphics[width=8.5cm]{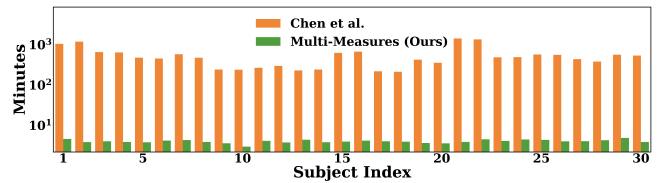}}
  \caption{Runtime comparison between the traditional method and our Multi-Measures approach (Ours).
}
  \label{fig:runtime}
\end{minipage}
\end{figure}

\subsection{Runtime Comparison}
\label{sec:Runtime}
\vspace{-0.2cm}
We report the runtime for 30 randomly selected subjects, where the horizontal axis denotes subjects and the vertical axis indicates execution time (Fig.~\ref{fig:runtime}). The traditional method requires the extremely long processing time, often taking several hours per subject, because the large amount of remaining gyral regions leads to time-consuming tree construction and pruning. In contrast, our Multi-Measures approach achieves a dramatically shorter and far more stable runtime, typically completing the entire pipeline within only a few seconds. This efficiency highlights the scalability of our framework and its suitability for large-scale population studies.

\section{Discussion and Conclusion}
\label{sec:conlusion}
In this work, we introduced GS-Net, a framework that integrates the brain’s two primary folding patterns, gyri and sulci, into a unified anatomical network. Unlike traditional region-based approaches, GS-Net represents fine-scale anatomical landmarks as network nodes, enabling more detailed and biologically meaningful characterization of cortical morphology. We evaluated GS-Net on multiple datasets totaling over 1,600 brain scans across diverse cohorts and age groups, demonstrating that GS-Net accurately captures individual folding patterns. Compared with baseline methods, GS-Net not only achieves higher accuracy but also significantly reduces computation time, providing a fast and robust tool for large-scale brain network studies.

\section{Compliance with ethical standards}
\label{sec:ethics}
This study used brain MRI data from public datasets, including Human Connectome Project (HCP) S1200 release, ADNI dataset, and dHCP datasets. As the data were openly accessible and governed by the respective licenses, no additional ethical approval was required.



\bibliographystyle{IEEEbib}
\bibliography{strings,refs}

\end{document}